\documentclass[aps,pre,twocolumn,showpacs,floatfix,superscriptaddress,nofootinbib]{revtex4}
\usepackage{graphicx}
\usepackage{amsmath}

\newcommand {\be} {\begin{equation}}
\newcommand {\ee} {\end{equation}}
\newcommand {\Be}{\begin{eqnarray*}}
\newcommand {\Ee} {\end{eqnarray*}}
\newcommand {\bey} {\begin{eqnarray}}
\newcommand {\eey} {\end{eqnarray}}
\newcommand{\bit}{\begin{itemize}}      
\newcommand{\eit}{\end{itemize}}
\newcommand{\bfl}{\begin{flusleft}}
\newcommand{\efl}{\end{flusleft}}
\newcommand{\bfr}{\begin{flushright}}
\newcommand{\ec}{\end{center}}
\newcommand{\ben}{\begin{enumerate}}    
\newcommand{\een}{\end{enumerate}}

\newcommand{\comment}[1]{}

\begin{document} 

\title{A microscopic mechanism for self-organized quasi periodicity in random networks of non linear oscillators}
\author{Raffaella Burioni}
\affiliation{Dipartimento di Fisica e Scienza della Terra,  Universit\`a di
Parma, via G.P. Usberti, 7/A - 43124, Parma, Italy}
\affiliation{INFN, Gruppo Collegato di Parma, via G.P. Usberti, 7/A - 43124, Parma, Italy} 
\author{Serena di Santo}
\affiliation{Dipartimento di Fisica e Scienza della Terra,  Universit\`a di Parma, via G.P. Usberti, 7/A - 43124, Parma, Italy}
\author{Matteo di Volo}
\affiliation{Dipartimento di Fisica e Scienza della Terra,  Universit\`a di Parma, via G.P. Usberti, 7/A - 43124, Parma, Italy}
\affiliation{Centro Interdipartimentale per lo Studio delle Dinamiche Complesse, via Sansone, 1 - 50019 Sesto Fiorentino, Italy}
\affiliation{INFN, Gruppo Collegato di Parma, via G.P. Usberti, 7/A - 43124, Parma, Italy}
\author{Alessandro Vezzani}
\affiliation{ S3, CNR Istituto di Nanoscienze, Via Campi, 213A - 41125 Modena, Italy}
\affiliation{Dipartimento di Fisica e Scienza della Terra,  Universit\`a di Parma, via G.P. Usberti, 7/A - 43124, Parma, Italy}
\begin{abstract}
Self-organized quasi periodicity is one of the most puzzling dynamical phases observed in systems of non linear coupled oscillators. The single dynamical units are not locked to the periodic mean field they produce, but they still feature a coherent behavior, through an unexplained complex form of correlation. We consider a class of leaky integrate-and-fire oscillators on random sparse and massive networks with dynamical synapses, featuring self-organized quasi periodicity, and we show how complex collective oscillations arise from constructive interference of microscopic dynamics. In particular, we find a simple quantitative relationship between two relevant microscopic dynamical time scales and the macroscopic time scale of the global signal. We show that the proposed relation is a general property of collective oscillations, common to all the partially synchronous dynamical phases analyzed. We argue that an analogous mechanism could be at the origin of similar network dynamics.
\end{abstract}
   

\maketitle


\section{Introduction}
The emergence of synchronized signals from networks of coupled oscillators is an intriguing and ubiquitous phenomenon, that has been observed in systems ranging from mechanical oscillators to neurons \cite{PRK,kura}. A large part of the dynamical units organize to produce an oscillating  global field, and the observed coherence indicates that a complex form of correlation is present between microscopic units. Of particular experimental interest is the coherent behavior occurring  in brain circuits, as in $\gamma$-oscillations or in  $\theta$-rhythm in the hippocampus \cite{busaki}. Firing neurons are indeed able to organize at a microscopic level to give rise to highly complex quasiperiodic signals \cite{luccioli,hansel,miguel,chimera,olmi, dearcangelis,strogatz,cowan,geisel,brunel}. 

The problem has been studied in details under multiple aspects \cite{carl,politi1,piko}, and the general mechanism, if one, that  organizes  the individual  microscopic motions  to produce macroscopic periodic oscillation is still not well understood. One of the most puzzling observed dynamical phase is self-organized periodicity \cite{piko,experiment,Vilfan}, where the single dynamical units are not locked to the periodic mean field they produce, but they still feature  a coherent behavior. There is no clear derivation of this mechanism, that arises in many different models of neuron dynamics and non linear oscillators coupled on different classes of  networks.

In this paper we present an analysis of self-organized quasi periodicity in two models of neurons dynamics with dynamical synapses on fully coupled networks, heterogeneous massive networks and sparse networks and we discuss a general mechanism leading to quasi-synchronous oscillations from the periods of microscopic dynamics. 

We consider  the $\alpha$ model on globally coupled networks, where the synaptic dynamics is implemented through a pulse of width $\alpha$ \cite{carl}, and the TUM model on heterogeneous massive networks with fixed in--degree distribution and on sparse random regular networks. In this case, the synapses are described by the short--term plasticity  mechanism that rules synaptic transmitters in the active, inactive and recovered state \cite{tso}. The $\alpha$ model on fully connected networks is an example of pure mean field dynamics without disorder or fluctuations, while in the TUM model on heterogeneous massive networks the mean field dynamics is characterized by topological disorder. In the case of sparse networks, the dynamics is not mean field and local fluctuations play a fundamental role. 

Although the dynamical phases of these models are different, we show that the underlying mechanism giving rise to a collective oscillation of a period that differs from that of the underlying microscopic dynamics is the same. We find a quantitative relation, holding in all regimes of partial synchronization, between the global oscillations and  two microscopic timescales:  these are a fast oscillation, related to the average inter-spike interval of neurons, and a slower oscillation in the average escape time from almost regular spiking. These two frequencies combine in a fixed and extremely simple relation to produce the periodicity of the global field. In particular, in the frequency space, the phases of local signals are correlated only for the frequency of the global field and randomly distributed at different frequencies. In this way the summation of quasi periodic local signals produces a result of the expected periodicity. 
This constructive interference appears to be a general property responsible for collective oscillations and common to all the partially synchronous dynamical phases analyzed here. An analogous mechanism could be at work in similar networks of coupled oscillators featuring self-organized quasi periodic signals.

The paper is organized as follows: in Section \ref{sec2} we introduce the leaky integrate-and-fire (LIF) neurons on directed networks and the two models for synaptic dynamics and we show that all the self-organized quasi periodic regimes considered feature different time scales in the global and microscopic dynamics. Section \ref{sec3} contains our main results: we show how complex collective oscillations arise from microscopic dynamics. In particular, taking first into account the simplest case of the $\alpha$ model on fully connected networks, we find a simple quantitative relationship between the microscopic dynamical time scales and the macroscopic ones. We then show that the same relation holds for the  TUM model on heterogeneous and sparse networks. Finally, we discuss our conclusions and perspectives in Section \ref{Conclusions}. In the Appendix \ref{app1} we describe in details the dynamical phases of the TUM model on sparse networks, providing a description of the main mechanism for its microscopic dynamics.

\section{LIF models with dynamical synapses and their dynamical phases}
\label{sec2}

In neural circuits the oscillations are often related to the presence of a balance between excitation and inhibition in the network \cite{inib}. Recently though, coherent quasi-synchronous signals have been detected in vivo also in networks of pure excitatory neurons \cite{bonifazi}. 
So we focus our analysis to this simpler case. In particular, the dynamics of each neuron $i$ in a network of $N$ nodes follows the equation:
\begin{equation}
\label{lif}
\dot v_i(t)= a -v_i(t) + \frac{g}{\langle k\rangle }\sum_j \epsilon_{i,j}f_j(t) ,
\end{equation}
where $v_i(t)$ is the rescaled membrane potential of neuron $i$, $a$ is the rescaled external current (equal for each neuron $i$), $g$ is the positive coupling strength, $f_i(t)$ is the synaptic field produced by neuron $i$ and $\epsilon_{i,j}$ is the  adjacency matrix of the directed uncorrelated network, whose entries are equal to $1$ if  neuron $j$ fires to neuron $i$ and $0$ otherwise. From the adjacency matrix, we define the in-degree of neuron $i$ as $k_i = \sum_j \epsilon_{i,j}$. Whenever the potential $v_i(t)$ reaches the threshold value $v_{th} = 1$, it is reset to $v_r = 0$, and a spike is
sent towards the postsynaptic neurons. 

We consider sparse and massive networks. In the first case the average in-degree  $\langle k\rangle$ remains finite in the thermodynamic limit $N\to\infty$, while, in massive networks,  $\langle k\rangle\sim N$. Notice that, with the renormalized coupling $1/\langle k\rangle$, the coupling term always remains finite  at increasing $N$. In all simulations we set $a=1.3$ so that the single neuron dynamics is in the firing regime. Similar results hold for any $a>1$, while in the single neuron dynamical phase with $a<1$ different dynamical regimes have been observed \cite{DL,lucciolihh}. 

We consider two different synaptic rules for $f_i(t)$. In the $\alpha$--model $f_i(t)$ is a pulse of width $1/\alpha$ \cite{carl} whose dynamics is defined by:

\begin{equation}
\label{alpha-field}
\ddot  f_i(t)+2\alpha \dot f_i(t)+ \alpha^2 f_i(t)= \alpha^2S_i(t).
\end{equation}
where, $S_i(t)=\sum_m \delta(t-t_{i}(m))$ is the spike train of neuron $i$  and $t_{i}(m)$ is the time when neuron $i$ fires its $m$-th spike. In the second case, $f_i(t)$ is described by the TUM model with  short--term--plasticity  \cite{tso}. For each excitatory neuron, the fraction of active, inactive and available resources are respectively $f_i(t)$, $z_i(t)$, $x_i(t)$, and the synaptic field produced by neuron $i$ is:
\begin{align}
\label{tum}
& \dot f_{i} = -\frac{f_{i}}{\tau_{\mathrm{in}}} +ux_{i}S_i\\
\label{contz}
& \dot z_{i} = \frac{f_{i}}{\tau_{\mathrm{in}}}  -   \frac{z_{i}}{\tau_{\mathrm{r}}} .
\end{align} 
with $x_i+f_i+z_i=1$. In all simulations we have fixed the synaptic parameters to phenomenological values, i.e. $\tau_{\mathrm{in}}=0.6$, $\tau_{\mathrm{r}}=79.8$ and $u=0.5$ \cite{tso,volman,fuchs}.  

We focus our analysis on three cases. First, we study the $\alpha$ model on  globally coupled networks, i.e. ordered  massive networks with $ \epsilon_{i,j}=1$ for all $i$ and $j$. In this model, neurons are identical and the interaction is mean field so that disorder and fluctuations do not affect the dynamics.  Then, we consider the TUM  model on massive networks with fixed in--degree distribution. These are defined, similarly to the configuration models \cite{conf}, introducing for each size $N$ a distribution of in-degree $P_N(k)$ so that $\langle k \rangle \sim N$. In these heterogeneous networks, the topological disorder in the in-degree plays a fundamental role in generating a complex dynamical phase  \cite{noi}. Indeed the same model on fully connected structures does not feature self-organized quasi periodicity \cite{DLLPT}, while quasi periodic regimes arise in the heterogeneous case. Finally, we study the TUM model on regular sparse networks, with no disorder in the in--degree, i.e. for each neuron $k_i=k$. This model features a non mean field dynamics, with local fluctuations of the synaptic field. Nevertheless, as shown in Appendix \ref{app1},  for large enough connectivity or for small enough couplings  this system presents a dynamical phase characterized by quasi-periodic oscillations. Although the dynamics of these three models displays very different features, we show that the mechanism underlying the emergence of complex collective behavior is the same. 

\begin{figure}
\centering
\includegraphics[width=8.5 cm]{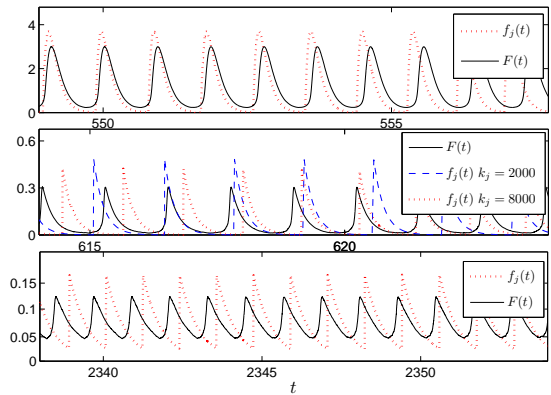}
\caption{(Color online) Black continuous lines are the average fields $F(t)$ while dashed and dotted lines are the local fields $f_i(t)$. Upper panel refers to the $\alpha$ model with $\alpha=10$, $g=0.4$ and $N=10^3$. The central panel refers to the TUM model with $g=21$ defined on a inhomogeneous massive network with  $N=10^4$ and $P_N(k)$ a Gaussian distribution  with $\langle k \rangle=0.7 N$ and $\sigma_k=.06 \langle k \rangle$. For these values, $k_{c1}\simeq 4500$ and $k_{c2}\simeq \langle k\rangle = 7000$. The lower panel describes the TUM model with $g=20$ on a sparse graph with $k=20$ and $N=500$. The same values of the parameters are used also in the following figures.}
\label{campi}
\end{figure}

The collective dynamics and its synchronization properties can be detected by analyzing the average field $F(t)=\sum_j f_j(t) /N$. This is an interesting quantity, that is also more easily experimentally recorded  than single neuron activities \cite{fmri1}. Furthermore, $F(t)$ encodes the complex behavior of the system, as the collective activity field is not a straightforward resultant of the fields $f_i(t)$. 

In the dynamical regime of partial synchronization, $F(t)$ can present periodic oscillations peaked at quasi--synchronous--events where a subgroup of neurons fire almost at the same time. However the single neuron dynamics is characterized by  time scales that differ from that of the global field $F(t)$. Fig. \ref{campi} shows that the local and the global fields display different frequencies. In the $\alpha$ model on globally coupled networks and in the TUM model on sparse networks, the period $T$ of $F(t)$ is larger than the average periodicity of $f_i(t)$. In these cases, the local dynamics is independent of the site $i$ and all neurons have the same frequency. Conversely, the TUM model on heterogeneous massive networks is characterized by topological inhomogeneity, and the period of local oscillations depends on the neuron in--degree $k_i$; neurons with $k_i\in[k_{c1},k_{c2}]$, are periodic of the same period of $F(t)$ and are called locked. Conversely neurons $i$ with $k_i>k_{c2}$ and with $k_i<k_{c1}$ are quasi--periodic with an average periodicity smaller and larger than $T$, respectively \cite{noi}. The comprehension of the mechanism yielding collective oscillations characterized by this complex periodicity pattern is of particular interest, as these oscillations are typically observed in experimental setups, e.g.  in mammalian brains, where such a coherent rhythmic behavior involves  different groups of neurons \cite{busaki}.

\section{Microscopic self--organization for collective dynamics}
\label{sec3}

In order to shed light on the microscopic self-organization of collective oscillations, we first consider the simplest case of the $\alpha$ model, where the physical time scales involved in the collective dynamics can be easily identified. Afterwards, we will consider both the sparse and massive TUM model, showing that the microscopic mechanism giving rise to collective motion is the same.

\subsection{Analysis in the direct space}

\begin{figure}
\centering
\includegraphics[width=8.5 cm]{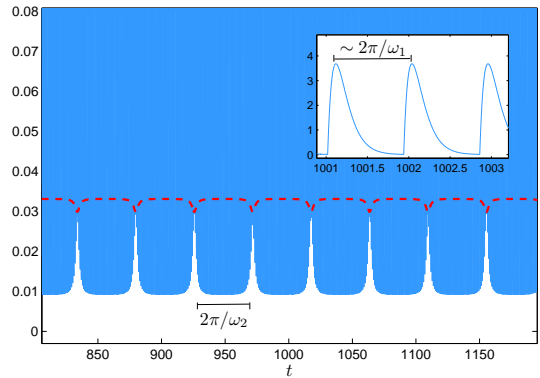}

\caption{(Color online) Microscopic field (continuous blue line) $f_i(t)$ for a neuron in a globally coupled network for the $\alpha$ model. In the inset we show a zoom with the fast oscillations of the microscopic dynamics. The dashed red curve is the sequence of $ISI_m$ at each $m$--th spike of the considered neuron that has been suitably rescaled in order to make it visible.}
\label{campimicro_alpha}
\end{figure}

In Fig. \ref{campimicro_alpha} we report the microscopic dynamics for the $\alpha$ model on globally coupled networks. The blue solid line represents the time evolution of the variable $f_i(t)$, common to all neurons $i$. As pointed out in \cite{carl}, the dynamics is quasi--periodic as it is characterized by two non--commensurable  time scales. Here we identify the physical origin of these time scales.  The first one is related to the fast oscillations reported in the inset.  These fast oscillations are not regular, as can be seen from the red curve representing the inter--spike--interval $ISI_m$ of single neurons $i$ at each m--th spiking time $t_i(m)$, i.e. $ ISI_m=t_i(m+1)-t_i(m)$. Thus, we define the frequency $\omega_1$ related to fast oscillations as an average of $ISI_m$, i.e.  $\omega_1=2\pi/\langle ISI\rangle$, where the average runs over the spiking times $t_i(m)$. The second time scale corresponds to regular slow periodic oscillations of $ISI_m$ as a function of the spiking time $t_i(m)$. These oscillations give rise to a straightforward definition of the frequency $\omega_2$.

Quasi-synchronous oscillations are present also in disordered models whose dynamics is characterized by local inhomogeneities. In these cases, the mechanism giving rise to collective motion is the same and it can be analyzed through the investigation of the same time scales suggested by the $\alpha$ model on all--to--all networks. However, the frequencies have to be defined taking into account the inhomogeneity of the system. In particular, for the massive TUM model with topological disorder, since the connectivity is not constant, neurons dynamics depends on their in--degree $k_i$.  The frequencies  $\omega_1(i)$ and $\omega_2(i)$ are defined from the average inter--spike interval and from the slow oscillations of $ISI_m(i)$ respectively; clearly, now, both frequencies depend on the in-degree of the neuron $k_i$. 
\begin{figure}
\centering
\includegraphics[width=8.5 cm]{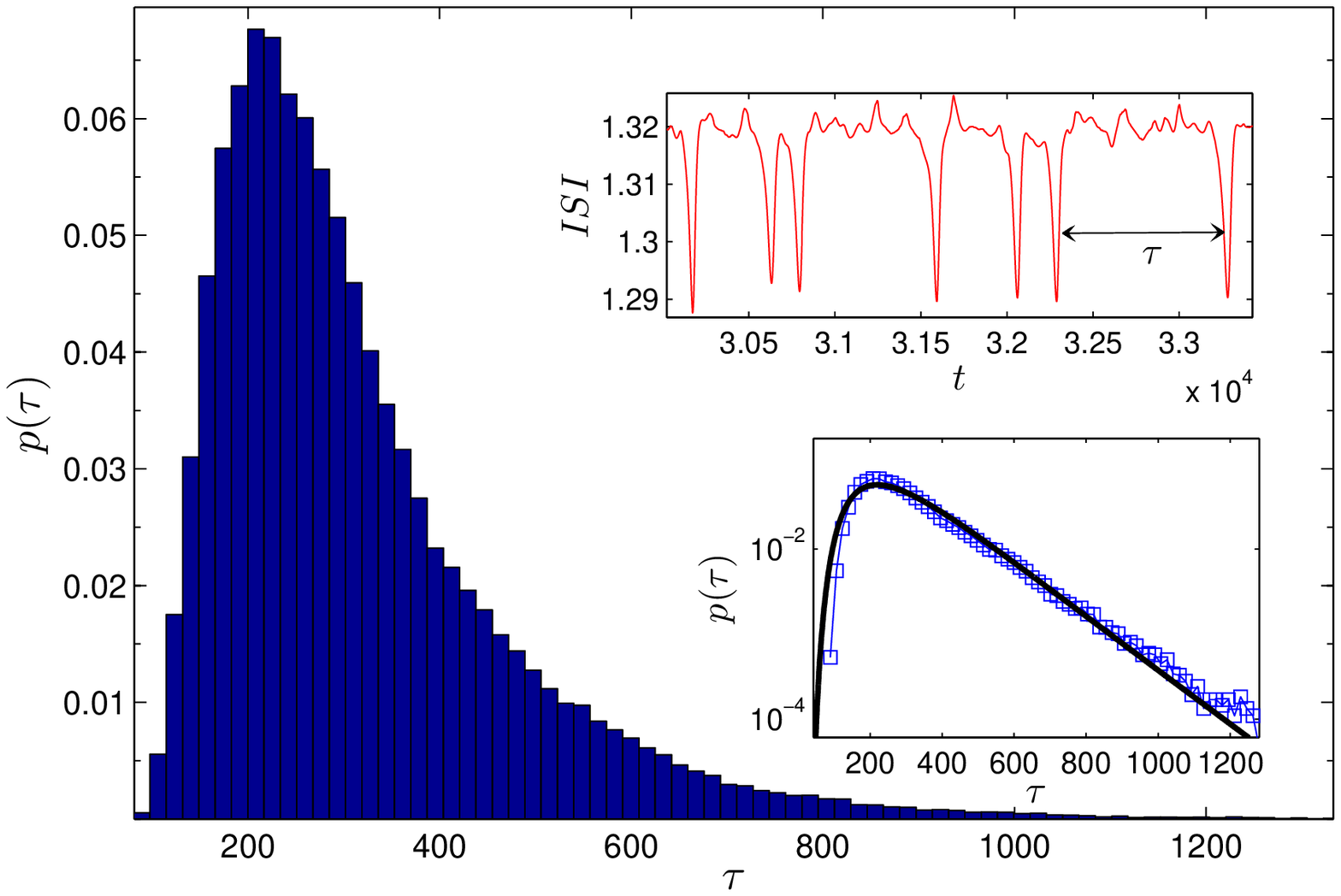}
\includegraphics[width=8.5 cm]{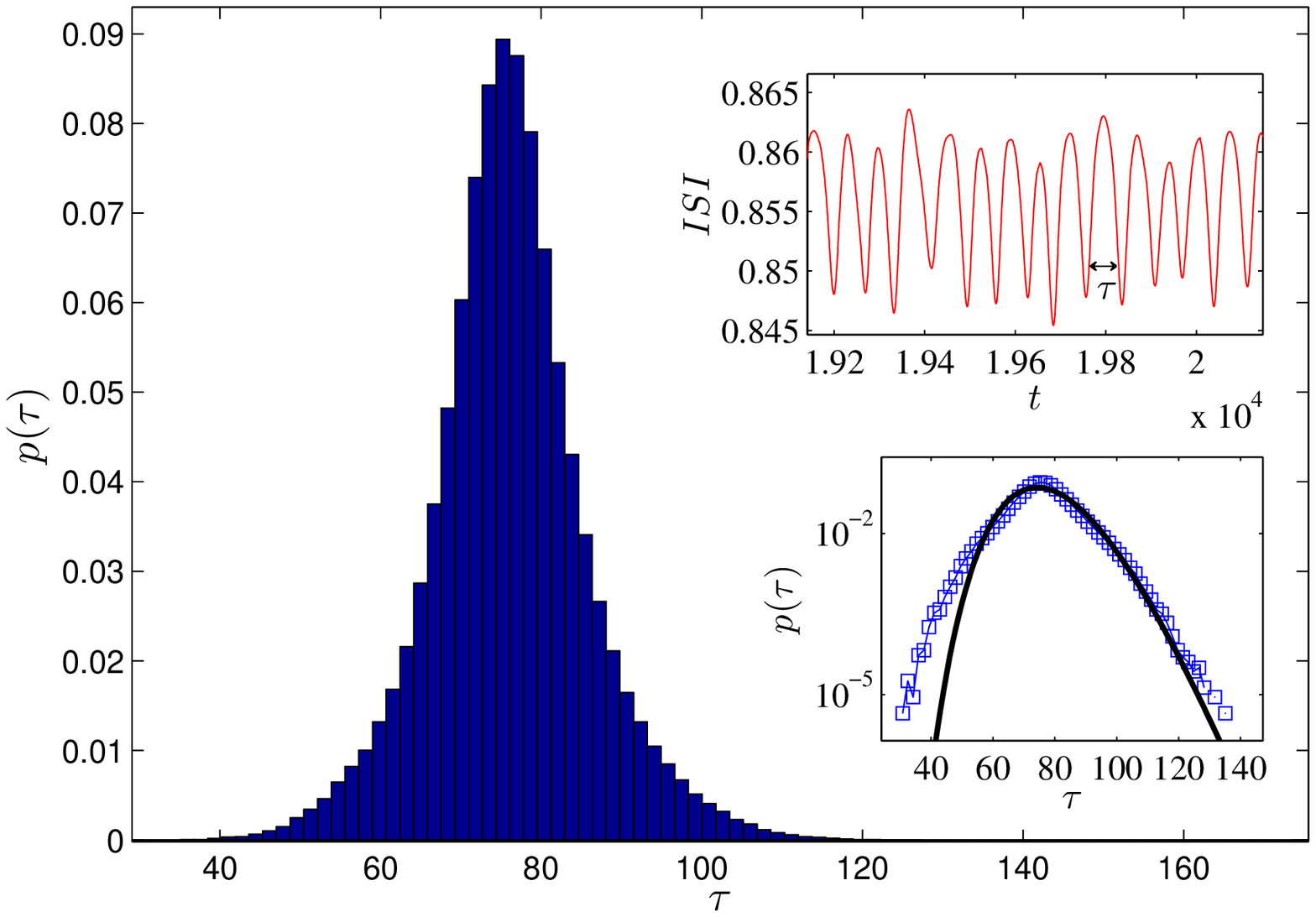}
\caption{(Color online) Escapes statistics for the TUM model in a sparse network of $N=500$ neurons, $k=20$ and $g=10$ (upper panel) and $g=60$ (lower panel).  In the upper inset of each panel we plot $ISI_m$ for a generic neuron. We define the time lapse between two escapes $\tau$ as the positive difference between consecutive peaks of $ISI_m$. The blue histogram is the probability distribution of these events obtained on a sample of $10^6$ escapes. In the lower inset of each panel we plot in log--lin scales the same distribution (squares) and a inverse Gaussian with the same average and variance of the data (continuous black line).
}
\label{fughe1}
\end{figure}

Finally, quasi-periodicity appears also in models defined on sparse networks where the dynamics of single units is ruled by the fluctuations of the received field; i.e. a mean field approach is not feasible. In Fig. \ref{fughe1} we show that in the TUM model on a regular sparse networks the escapes from periodic firing that correspond to the frequency $\omega_2$ are not regular and their statistics is well fitted by an inverse Gaussian characterized by an exponential tail. As shown in Appendix \ref{app1}, the escapes statistics is induced by the fluctuations of the  field received by single neurons, with an Arrhenius behavior. Accordingly the frequency $\omega_2$ can be defined only statistically by averaging over the distribution of the escapes $\tau$, i.e. $\omega_2=2\pi/\langle \tau \rangle$. Nevertheless, even in this complex case dominated by fluctuations, we can introduce the slow and the fast frequencies $\omega_2$ and  $\omega_1=2\pi/\langle ISI \rangle$.

\begin{figure}
\centering
\includegraphics[width=8.5 cm]{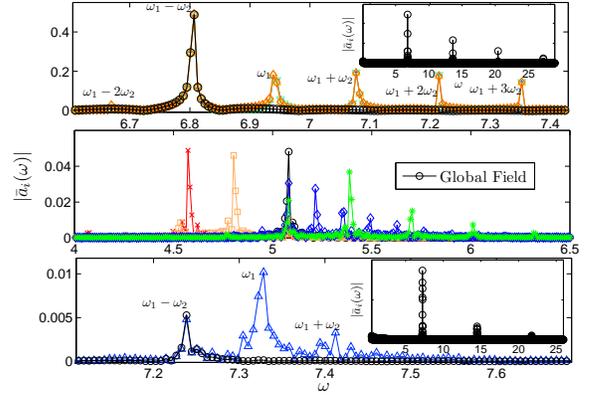}
\caption{(Color online) Modulus of the Fourier Transform of the macroscopic fields $F(t)$ (black circles) and of the local fields relevant to different neurons $f_i(t)$ (other symbols). Upper, middle and lower panels refer to $\alpha$ model on fully connected graph, TUM on heterogeneous massive and sparse random networks respectively. }
\label{fourier}
\end{figure} 

\subsection{Analysis in the frequency space}

In Fig. \ref{fourier} we plot the Fourier transform of $F(t)$ and  $f_i(t)$. The global field $F(t)$ is characterized by a  single frequency $\Omega$ with its multiples (see the insets). Local fields $f_i(t)$ display peaks at integer combinations of $\omega_1$ and $\omega_2$ i.e.
\begin{equation}
f_i(t)=\sum_{n_1,n_2\in Z}\bar a_i(\omega) e^{I\phi_i(\omega)}e^{I(n_1\omega_1+n_2\omega_2)t},
\label{fourier1}
\end{equation}
where $\bar a_i(\omega)$ and $\phi_i(\omega)$ are respectively the modulus and the phase corresponding to the harmonic $\omega=n_1\omega_1+n_2\omega_2$. Notice that $\omega_1$ and $\omega_2$ depend on $i$ for the TUM model on heterogeneous massive networks. Fig. \ref{fourier} shows that the macroscopic frequency $\Omega$ sets in a simple combination of the microscopic frequencies; in particular for the $\alpha$ model and for the TUM model on sparse network $\Omega=\omega_1-\omega_2$. For the TUM model on heterogeneous massive networks, we have that for locked neurons only a single frequency $\omega=\Omega$ is present while for each unlocked neuron $\Omega\simeq \omega_1(i) \pm \omega_2(i)$ where the $+$ applies if $k_i<k_{c1}$ and $-$ if $k_i>k_{c2}$.

\begin{figure}
\centering
\includegraphics[width=8.5 cm]{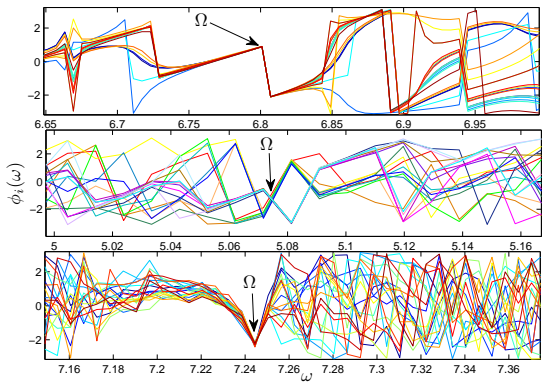}
\caption{(Color online) Phases $\phi_i(t)$ of the Fourier Transform of the local fields.  Upper, middle and lower panels refer to $\alpha$ model on fully connected graph, TUM on heterogeneous massive and sparse random networks respectively. $\Omega$ is the characteristic frequency of the global field.}
\label{fasi}
\end{figure}

Fig. \ref{fasi} shows that collective oscillations of the global field are a consequence of the distribution of the phases $\phi_i(\omega)$ as a function of $i$ and $\omega$.  In particular, the phases $\phi_i(\omega)$ of different neurons are uncorrelated except for the values $\omega=n \Omega$ corresponding to the frequencies of the global field, i.e  $\Omega=\omega_1-\omega_2$, for the alpha model and the TUM model on sparse networks and   $\Omega=\omega_1 \pm \omega_2$ for the TUM model on heterogeneous structures. This implies that the global field is characterized only by the frequency $\Omega$.

\begin{figure}
\centering
\includegraphics[width=8.5 cm]{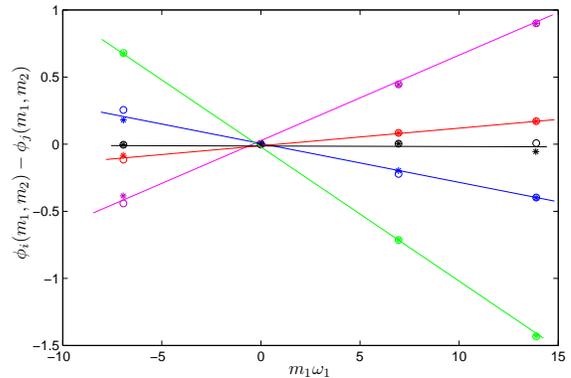}
\caption{(Color online) Phase differences $\phi_i(m_1,m_2)-\phi_j(m_1,m_2)$ for different couples of neurons as a function of $\omega_1 m_1$. Lines connect points related to the same couple. Asterisks and circles represent peaks in the Fourier transform in Eq. (\ref{fourier2}) having the same value of $m_2$ but different value of $m_1$.}
\label{diff_fasi_alpha}
\end{figure}

For the globally coupled $\alpha$ model the Fourier transform can be 
discussed in details. 
We notice that the modulus $\bar a_i(\omega)$ is independent of the neuron $i$ and Eq.(\ref{fourier1}) can be rephrased as:
\begin{equation}
f_i(t)=\sum_{m_1,m_2\in Z}\bar a(m_1,m_2) e^{I\phi_i(m_1,m_2)}e^{I(m_1\omega_1+m_2\Omega)t},
\label{fourier2}
\end{equation}
Fig. \ref{diff_fasi_alpha} shows that for any values of $m_1$ and $m_2$ the phase difference between two neurons $i$ and $j$ is independent of $m_2$ and it is proportional to $m_1\omega_1$. Therefore, one has $\phi_i(m_1,m_2)=m_1 \omega_1 t_i+\psi_{m_1,m_2}$ where  $t_i$ and $\psi_{m_1,m_2}$ are two constant. The first depends only on the neuron $i$ and the latter is independent of the neuron. It is then possible to perform the sum in the Eq. (\ref{fourier2}), obtaining $f_i(t)=g(\omega_1(t-t_i),\Omega t) $ where $g(\cdot,\cdot)$ is a function periodic in both arguments and $t_i$ can be interpreted as a shift in the initial time due to different initial conditions. In the TUM massive model the same behavior is observed for every neuron sharing the same connectivity $k_i$, but in general the shape of the function $g$ and $\omega_1$ depend on $k_i$. In the sparse case this scenario cannot be observed as the presence of large fluctuations induces large fluctuations also in the phases of the Fourier transform.

\begin{figure}
\centering
\includegraphics[width=8.5 cm]{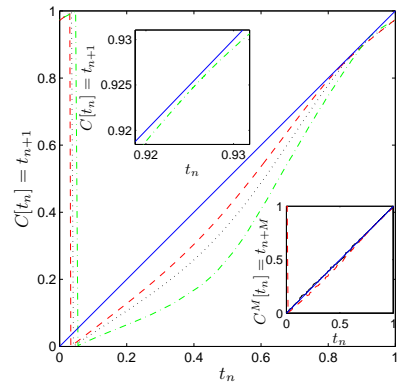}
\caption{(Color online) Map $C[t]$ for the globally coupled $\alpha$ model. Continuous (blue) line is the bisector of the square and dashed (red), dotted (black) and dash-dotted (green) lines are the maps of single neurons for $\alpha = 9$, $\alpha =10$ and $\alpha=15$ respectively. The zoom in the upper inset clarifies that the map never intersects the bisector of the square.  In the lower inset we plot the map $C^M$, i.e. $t_{n+M}=C^M[t_n]$ (see text for the definition of $M$). }
\label{map}
\end{figure} 

\subsection{A heuristic interpretation and the robustness of the result}

A heuristic interpretation of this phenomenon can be given in the direct space by studying the microscopic dynamics through a Poincar\'e map.
At each $n$--th peak of the global field at times $nT$ we report the positive firing time delay with respect to the peak for a given neuron, and we call it $t_n$. Thus we obtain a map $C$ such that $t_{n+1}=C[t_n]$; Fig. \ref{map} reports this map for the $\alpha$ model.
 Despite the neuron spends most of the time firing at the same delay with respect to the peak of $F(t)$ (where the curve is almost tangent to the bisector), the dynamics does not have a fixed point (see the upper inset).  Therefore, the delay between the firing time and the peak of $F(t)$ reduces at each firing event, until it becomes negative. In this case the positive delay has to be measured taking into account the previous peak of $F(t)$  giving rise to the discontinuity of the map in Fig. \ref{map}.
In this perspective the longer time $2\pi/\omega_2$ can be interpreted as the time that the neuron takes before coming back firing with the initial delay with respect to the global field. Suppose that in this time interval the neuron has fired $M$ times: then we should have $\omega_1= M\omega_2$. Indeed the lower inset of Fig. \ref{map} shows that the $M$--th iterate of the map is almost the identity ($M$ is the nearest integer to $\omega_1/\omega_2$ and the frequencies have been taken from the data of Fig. \ref{campimicro_alpha}).
Furthermore, as the neuron anticipates its firing, the global field has performed $M-1$ oscillations in the same time, then $\Omega/\omega_2= M-1$ and  $\Omega = \omega_1-\omega_2$. 
\begin{figure}
\centering
\includegraphics[width=8.5 cm]{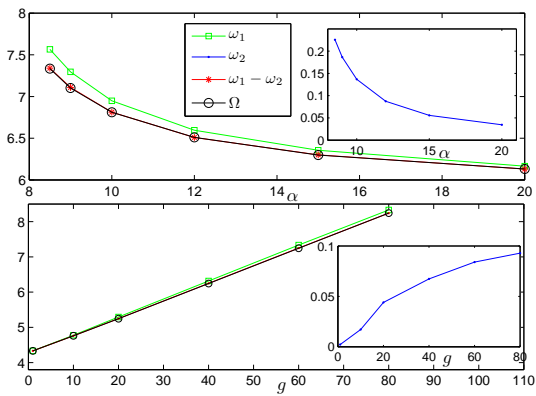}
\caption{(Color online) Upper panel describes the behavior of the characteristic frequencies of the fully connected $\alpha$ model as a function of $\alpha$. In the inset we the plot shape of $\omega_2$ as a function of $\alpha$. In the lower panel the same frequencies for the TUM model on a sparse graph. The  inset shows the shape of $\omega_2$ as a function of $g$.}
\label{generality}
\end{figure} 

For the TUM model on heterogeneous massive networks, similar results are obtained introducing a map for each class of neurons depending on their connectivity. For locked neurons, the map has a fixed point and $\omega_2$ cannot be defined. Neurons with  $k_i>k_{c2}$ display the same dynamics of the $\alpha$ model, while neurons with $k_i<k_{c1}$ are unlocked but they postpone their firing ($\omega_1<\Omega$) and organize so that  $\Omega\simeq \omega_1+\omega_2$, since now, after $M$ spikes of the neuron, the global field has performed $M+1$ oscillations. Finally, for the TUM model on sparse networks, due to fluctuations, the map $C[t_n]$ cannot be defined. However an analogous mechanism of delay between local and global fields is present, leading to the same relation between frequencies.

Finally, we remark that the scenario holds in the whole parameter space where collective oscillations are observed. For the $\alpha$ model Fig. \ref{generality} shows the dependence of the frequencies $\omega_1$, $\omega_2$ and $\Omega$ as a function of $\alpha$.  In the inset, $\omega_2$ tends to zero for  $\alpha \to \infty$, i.e. when coupled through $\delta$--like pulses, and neurons seem to have the same periodicity of the global field. For the TUM model on sparse networks Fig. \ref{generality} shows the characteristic frequencies as a function of the coupling: the relationship between macroscopic and microscopic time scales remains unchanged. In this case $\omega_2$ tends to zero for small $g$, showing a full synchronization for vanishing coupling.

\section{Conclusions}
\label{Conclusions}
Self organized quasi-periodicity is an interesting and widely observed dynamical phase, generated by correlations in microscopic dynamics of single units, that cooperate to produce a coherent global signal. In this paper we have considered a class of leaky integrate-and-fire oscillators on massive and sparse random networks with dynamical synapses, to investigate the mechanism that, from microscopic time scales, gives rise to a different periodicity in the global signal. Both for the $\alpha$ model on fully connected topology and for the TUM model on massive and sparse networks we found a quantitative simple relationship between two microscopic dynamical time scales and the macroscopic one. In particular, the two relevant time scales are the average inter-spike interval $\langle ISI \rangle$ and the average escape time $\langle \tau \rangle$ from almost regular spiking, that combine in a sum or in a difference. This simple relation holds in the whole regime of partial synchronization analyzed here and it appears to be a general property of collective oscillations. In this perspective, our result could pave the way to an analysis of cross-correlations between microscopic motions in systems of coupled non linear oscillators.

\appendix
\section{The TUM model on sparse random networks}
\label{app1}
In this Appendix, we provide some details of the dynamical phases of the TUM model on sparse directed random networks.

\begin{figure}
\centering
\includegraphics[width=8.5 cm]{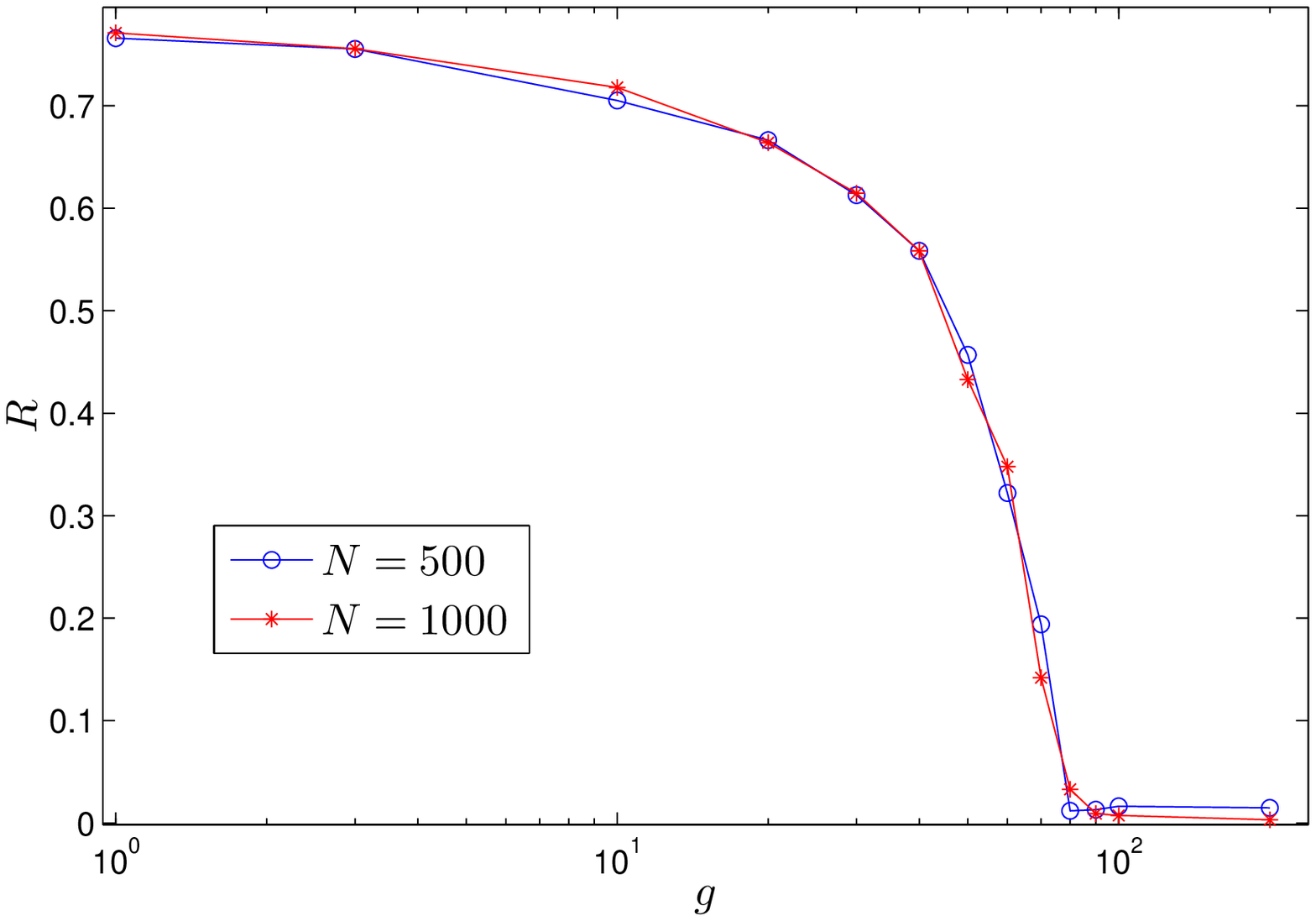}
\includegraphics[width=8.5 cm]{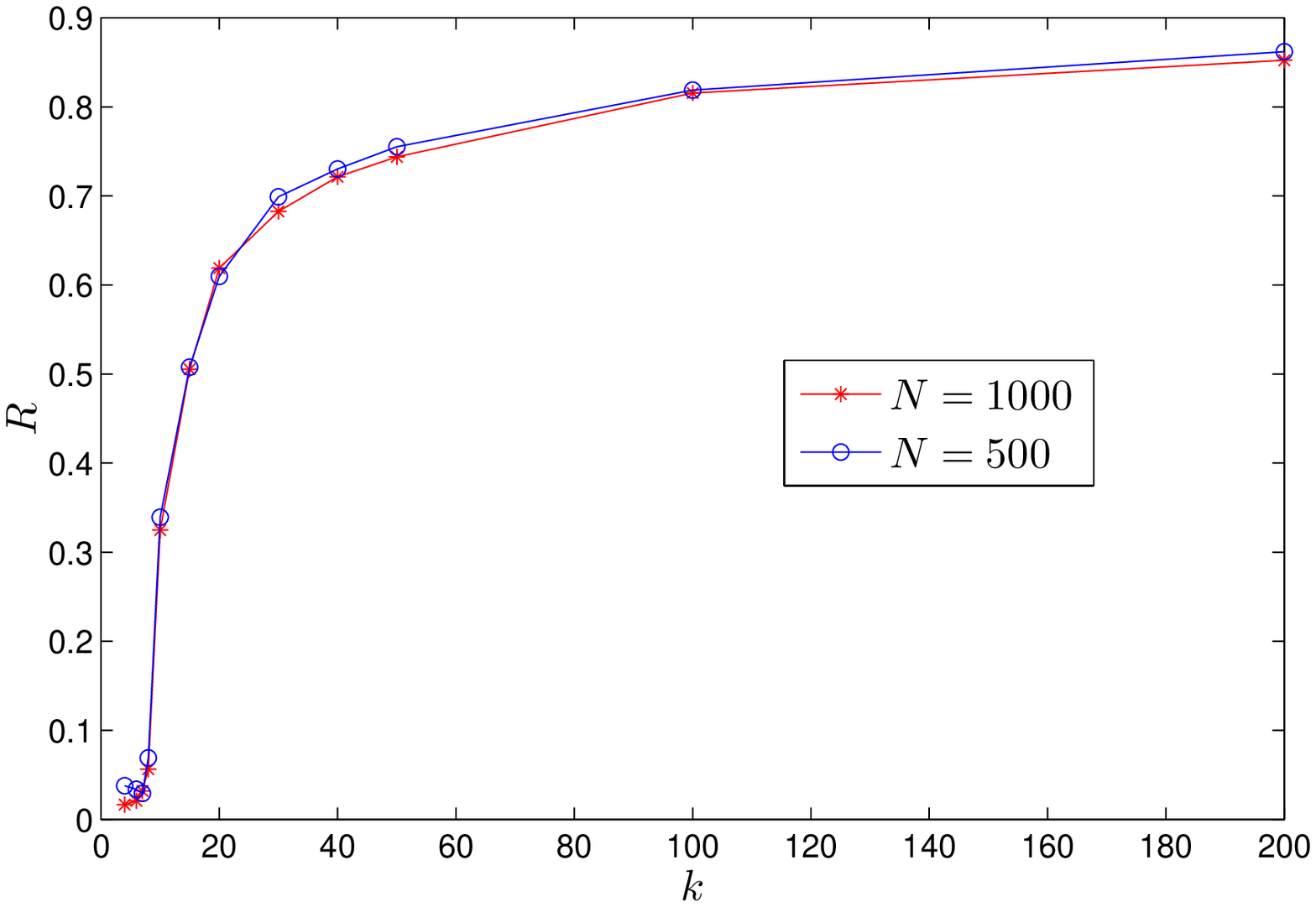}
\caption{(Color online) Kuramoto parameter (see text) in function of connectivity $k$ ($g=30$) and coupling $g$ ($k=20$) for sparse networks with different sizes $N=500$ and $N=1000$.}
\label{kura}
\end{figure}

The dynamics  of the model is described by equations (1), (2), (3) and (4), where now the adjacency matrix $\epsilon_{i,j}$ in equation (1) describes a directed random networks with $N$ sites and constant and finite in--degree $k_i=k$. Each neuron displays the same dynamical equation and fluctuations are induced by the fact that each neuron is driven by the field generated by different neighboring sites. To analyze the synchronization properties of the network, we plot in Fig. \ref{kura} the Kuramoto parameter $R$ as a  function of the coupling $g$ and of the degree $k$ \cite{kura}. 
We define at time $t$ for each neuron $i$ its phase $\theta_i(t)= (t-t_i(m))/ISI_{m}$, where $t\in[t_i(m),t_i(m+1)]$ so that:
\begin{equation}
R=\big \langle \Bigg| \frac{\sum_ie^{I\theta_i(t)}}{N} \Bigg| \big \rangle_t,
\end{equation}
where $I$ is the complex unit,  the average runs over time $t$ and $|...|$ is the usual modulus for complex numbers. If the neurons are synchronized, i.e. they fire at the same time, $R=1$, while in the completely asynchronous regime $R=0$.

In analogy with the analysis reported in \cite{luccioli}, where the same network structure is considered in the $\alpha$ model case, increasing the degree makes neurons more synchronized. In particular, in the case of the lower panel of Fig. \ref{kura} the degree  plays the role of a control parameter bringing the network from a synchronous to an asynchronous state.  We observe a similar scenario as a  function of the coupling but now increasing the coupling yields a decrease in the synchronization level. This scenario appears in different neural network models, and can be interpreted  by considering that a larger coupling  amplifies the fluctuations of the field relevant to each neuron, inhibiting the possibility of mutual synchronization \cite{hansel}.  In particular, we have observed the same behavior also for the $\alpha $ model and in massive networks.

By looking at the microscopic features of single neurons, we observe that their $ISI_m$ yields the same dynamics.  Furthermore, in the regimes where oscillations arise, firing of neurons is not strictly periodic and the inter--spike interval $ISI_m$  is time dependent. In the inset of Fig. \ref{fughe1},  the red curve shows the dynamics of $ISI_m$ for a representative neuron. Far from the transition point (upper panel) we see a dynamics characterized by an average $ISI$ and some escapes, i.e. the neuron mostly fires with almost the same period and then escapes from this regime firing with a higher frequency. In the upper panel of Fig. \ref{fughe1} we plot the probability distribution of the time lapses between two consecutive escapes $\tau$ for $g=10$. 
Near the transition point, one observes a similar scenario and the distribution of escapes looks more symmetric.  Interestingly, both curves are well fitted by an inverse Gaussian with an exponential tail, a distribution that is typical of first passage properties of random motion in presence of drift \cite{lucciolihh}. In particular, the drift mechanism should here be related to the fact that, during the escapes, the $ISI$ is always smaller than its characteristic value, i.e. the neuron tends to anticipate its firing giving rise to a preferential direction in the $ISI_m$ dynamics.

\begin{figure}
\centering
\includegraphics[width=8.5 cm]{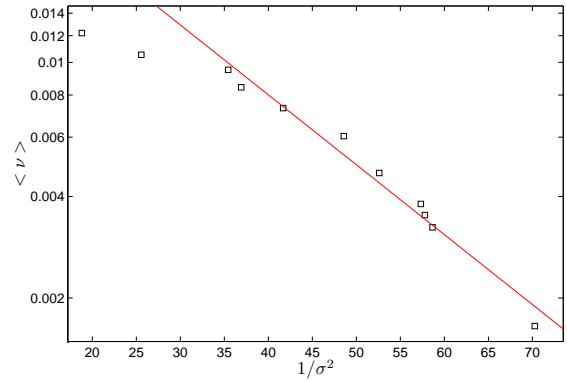}
\caption{(Color online) Neurons average escape frequency $\langle \nu \rangle  \equiv 1/ \langle \tau \rangle$ as a function of the noise variance on the received field. Each value has been obtained by averaging over $5$ single neuron dynamics. The continuous line is the curve $\langle \nu \rangle \sim \mathrm{exp}(-W/\sigma^2)$, with $W=0.05$. The in--degree is $k=20$ while the coupling $g$ changes in order to obtain different values of $\sigma^2$.  }
\label{fughe_sigma}
\end{figure}

To better quantify the distribution of escape times, we notice that the main mechanism of escapes can be seen as a trapping dynamics: the neuron is "trapped" to fire with the same $ISI_m$ for a certain time, and then escapes from the trap, firing with a different $ISI_m$. We argue that the escape mechanism could be driven by fluctuations in the local field felt by the neurons, with an Arrhenius behavior, and we expect the fluctuations to become larger while approaching the transition point. To check this picture, we measure the fluctuation of the local field by calculating its variance $\sigma$. i.e the difference at each time between the global field, averaged over all neurons, and the local field felt by a single neuron. Then we plot the average escape frequency $\langle \nu \rangle  \equiv 1/ \langle \tau \rangle$ as a function of the variance of the received field. Interestingly, as shown in Fig. \ref{fughe_sigma}, our numerical results are in very good agreement with a law of the type $\langle \nu \rangle \sim \mathrm{exp}(-W/\sigma^2)$, providing a  very good evidence of the proposed trapping mechanism with a barrier $W$, at least until the variance is smaller than $W$.

From our analysis, it is clear that one can define two main time scales in the dynamics. The first one is the average $\langle ISI \rangle$ (the average runs over the different firing events $m$) and the second one is the average escape time $\langle \tau \rangle$, i.e. the average of the distributions we plot in Fig. \ref{fughe1}. In a model ruled by fluctuation, this average timescales are the basis for the definition of the microscopic characteristic frequencies characterizing the dynamics of quasi-synchronous events, i.e. $\omega_1=2\pi/\langle \tau\rangle$ and $\omega_2=2\pi/\langle ISI\rangle$.
As a final remark, we point out that the same dynamical scenario is observed in random regular networks, where the output degree is also constant. This indicates that the fluctuations giving rise to this regime have a complex origin and are not only caused by topological disorder in the out-degree.

\end{document}